\documentclass{jetpl}
\twocolumn

\usepackage{amsmath,amssymb,epsfig,color,pstricks,bm,graphics,alltt}

\usepackage[colorlinks,plainpages=false,linkcolor=black,urlcolor=blue,citecolor=black,pdfpagemode=UseNone,pdfstartview=FitBH]{hyperref}

\definecolor{MyDarkBlue}{rgb}{0,  0.3,  0.9}
\definecolor{MyDarkBlack}{rgb}{0,  0,  0}

\begin{document}

\lat

\title{Disorder Effects in BCS-BEC Crossover Region of Attractive Hubbard
Model.}


\rtitle{Disorder Effects in BCS-BEC Crossover Region}

\sodtitle{Disorder Effects in BCS-BEC Crossover Region of Attractive Hubbard
Model.}

\author{$^a$E.\ Z.\ Kuchinskii\thanks{E-mail: kuchinsk@iep.uran.ru},
$^a$N.\ A.\ Kuleeva\thanks{E-mail: strigina@iep.uran.ru},
$^{a,b}$M.\ V.\ Sadovskii\thanks{E-mail: sadovski@iep.uran.ru}}

\rauthor{E.\ Z.\ Kuchinskii, N.\ A.\ Kuleeva, M.\ V.\ Sadovskii}

\sodauthor{E.\ Z.\ Kuchinskii, N.\ A.\ Kuleeva, M.\ V.\ Sadovskii}

\sodauthor{E.\ Z.\ Kuchinskii, N.\ A.\ Kuleeva, M.\ V.\ Sadovskii}

\address{$^a$Institute for Electrophysics, Russian Academy of Sciences, 
Ural Branch, Amundsen str. 106,  Ekaterinburg, 620016\\
$^b$Institute for Metal Physics, Russian Academy of Sciences, Ural Branch, 
S. Kovalevskaya str. 18, Ekaterinburg, 620990}

\abstract{We study the disorder effects upon superconducting transition
temperature $T_c$ and the number of local pairs in attractive Hubbard model within the
combined Nozieres --- Schmitt-Rink and DMFT+$\Sigma$ approximations.
We analyze the wide range of attractive interaction $U$, from the weak coupling 
region, where instability of the normal phase and superconductivity are well
described by BCS model, to the limit of strong coupling, where superconducting
transition is determined by Bose--Einstein condensation of compact Cooper pairs,
forming at temperatures much higher than superconducting transition temperature.
It is shown that disorder can either suppress $T_c$ in the weak coupling limit, or
significantly enhance $T_c$ in the case of strong coupling. However, in all
cases we actually prove the validity of generalized Anderson theorem, so that
all changes of $T_c$ are related to change of the effective bandwidth due
to disorder. Similarly, disorder effects on the number of local pairs are only
due to these band-widening effects.}


\maketitle

\section{Introduction}

The problem of superconductivity in the limit of strong coupling has attracted
theorists for rather long time \cite{Leggett}. The significant progress in this
field was achieved by Nozieres and Schmitt-Rink \cite{NS}, who proposed an
effective method to study the crossover from weak coupling BCS behavior to
Bose-Einstein condensation (BEC) in strong coupling region. In recent years
the progress of experimental studies of ultracold quantum gases in magnetic 
and optical dipole traps, as well as in optical lattices, allowing controllable
change of density and interaction parameters (see reviews \cite{BEC1,BEC2}) 
has also increased the interest to studies of BCS--BEC crossover. One of
the simplest models allowing the study of BCS--BEC crossover is the Hubbard
model with attractive interaction.

The most effective theoretical method to study strongly correlated systems
both in the case of repulsive interactions and in the case of attraction 
(including the region of BCS--BEC crossover) is the dynamical mean-field theory
(DMFT) \cite{pruschke,georges96,Vollh10}. Within the framework of DMFT the
attractive Hubbard model has already been studied in the number of papers
\cite{Keller01,Toschi04,Bauer09,Koga11}. However, there are only few 
works devoted to the studies of disorder effects on the properties of
normal and superconducting phases in this model. Qualitatively the influence
of disorder on the superconducting critical temperature $T_c$ in the region of 
BCS--BEC crossover was studied in Ref. \cite{PosSad}. Diagrammatic approach to
the analysis of disorder effects upon $T_c$ and normal phase properties in
the crossover region was developed in Ref. \cite{PalStr}. Recently we have
studied \cite{JETP14} the disorder influence on single-particle properties
and optical conductivity in disordered attractive Hubbard model within our
general DMFT+$\Sigma$ approach \cite{UFN12}, which is especially 
convenient to take into account different additional interactions like 
scattering by short-range order parameter fluctuations 
\cite{JTL05,PRB05,FNT06,PRB07},
disorder \cite{HubDis,HubDis2} or electron-phonon interaction \cite{e_ph_DMFT}. 
In this paper we use DMFT+$\Sigma$ approach combined with 
Nozieres --- Schmitt-Rink approximation \cite{NS} to
study the influence of disorder upon superconducting transition temperature
$T_c$ and the number of local pairs in attractive Hubbard model for the wide
range of interaction parameter $U$, including the BCS-BEC crossover region.

\section{Basics of Nozieres --- Schmitt-Rink and DMFT+$\Sigma$ approaches.}

We shall consider disordered attractive Hubbard model with the Hamiltonian:
\begin{equation}
H=-t\sum_{\langle ij\rangle \sigma }a_{i\sigma }^{\dagger }a_{j\sigma
}+\sum_{i\sigma }\epsilon _{i}n_{i\sigma }-U\sum_{i}n_{i\uparrow
}n_{i\downarrow },  
\label{And_Hubb}
\end{equation}
where $t>0$ is the transfer integral between nearest neighbors on the lattice, 
$U$ is Hubbard onsite attraction, 
$n_{i\sigma }=a_{i\sigma }^{\dagger }a_{i\sigma }^{{\phantom{\dagger}}}$ 
is electron number operator on the lattice site,  
$a_{i\sigma }$ ($a_{i\sigma }^{\dagger}$) 
is electron annihilation (creation) operator with spin projection $\sigma$ and 
local energies $\epsilon _{i}$ are assumed to be independent random variables 
on different lattice cites. To simplify diagrammatic analysis we assume the 
Gaussian distribution for $\epsilon _{i}$:
\begin{equation}
\mathcal{P}(\epsilon _{i})=\frac{1}{\sqrt{2\pi}\Delta}\exp\left(
-\frac{\epsilon_{i}^2}{2\Delta^2}
\right)
\label{Gauss}
\end{equation}
Parameter $\Delta$ here is the measure of disorder and the Gaussian
random field with short-ranged (``white-noise'') correlations is equivalent to
the usual ``impurity'' scattering, leading the the standard diagram technique
for the averaged Green's functions \cite{Diagr}.

In the following we shall consider the model system with ``bare''
semi-elliptic density of states (per elementary lattice cell and one spin 
projection) given by:
\begin{equation}
N_0(\varepsilon)=\frac{2}{\pi D^2}\sqrt{D^2-\varepsilon^2}
\label{DOS13}
\end{equation}
so that the bandwidth is $W=2D$. All calculations below were made for the case
of quarter-filled band (electron density per cite n=0.5). 

In the absence of disorder superconducting transition temperature  was 
analyzed in this model in a number of papers \cite{Keller01,Toschi04,Koga11} 
both from the condition of Cooper instability of the normal phase \cite{Keller01} 
(divergence of Cooper susceptibility) and also from the condition of
superconducting order parameter becoming zero at $T_c$ \cite{Toschi04,Koga11}. 
In Ref.  \cite{JETP14} we have determined this critical temperature from the
condition of instability of the normal phase, as reflected in specific
instability of DMFT iteration procedure. The results obtained in this way
in fact just coincide with the results of Refs. \cite{Keller01,Toschi04,Koga11}.

The essence of Nozieres -- Schmitt-Rink approach \cite{NS} to calculation 
of $T_c$ in the wide region of coupling strengths $U$, providing an effective 
interpolation from weak to strong coupling (including the BCS--BEC crossover 
region) is to solve the BCS equation for transition temperature:
\begin{equation}
1=\frac{|U|}{2}\int_{-\infty}^{\infty}d\varepsilon N_0(\varepsilon)\frac{th\frac{\varepsilon -\mu}{2T_c}}{\varepsilon -\mu} ,
\label{BCS}
\end{equation}
jointly with an equation for chemical potential (implicitly determined by the
band-filling), which actually controls $T_c$ in strong coupling BEC region. 
In Ref. \cite{JETP14} we have shown that such
calculations, with an equation for chemical potential solved via DMFT, produce
the dependence  $T_c$ on $U$, which is in almost quantitative agreement with 
results obtained via much more time-consuming exact DMFT calculations.
This is rather surprising, because of neglect of all vertex corrections
due to $U$ (ladder approximation) in Eq. (\ref{BCS}), especially in the region 
of large $U$. Apparently this signifies rather small role of these vertex 
corrections (fluctuation effects) for BCS-like instability both in crossover and
strong coupling regions. However, in calculations of chemical potential
$\mu$ (controlling $T_c$ for large $U$) these corrections are quite important and only their correct account 
within DMFT allows us to obtain the correct behavior of $T_c$ in the limit of
large $U$.

This allows us to calculate $T_c$ for the case of disordered attractive 
Hubbard model using the same approach. Actually, we shall solve Eq. (\ref{BCS}),
from which all corrections due to disorder scattering just drop out, except 
those leading to disorder widening of the density of states 
\cite{SCLoc} (replacing 
$N_0(\varepsilon)$ in Eq. (\ref{BCS}) by disorder renormalized density of
states), jointly with an equation for chemical potential, obtained via 
DMFT+$\Sigma$ procedure
\cite{UFN12}, which takes into contributions due to disorder, producing the 
the chemical potential for different values of $U$ and disorder $\Delta$.

This generalized  DMFT+$\Sigma$ approach \cite{JTL05,PRB05,FNT06,UFN12} 
supplies the standard dynamical mean-field theory (DMFT) 
\cite{pruschke,georges96,Vollh10}  with an additional (``external'')
self-energy $\Sigma_{\bf p}(\varepsilon)$ (which can in general be momentum
dependent), taking into account any possible interaction outside the DMFT,
which gives an effective calculation method for either single-particle or
two-particle properties \cite{PRB07,HubDis}. The success of this generalized
approach is connected with the choice of the single-particle Green's function
in the following form:
\begin{equation}
G(\varepsilon,{\bf p})=\frac{1}{\varepsilon+\mu-\varepsilon({\bf p})-\Sigma(\varepsilon)
-\Sigma_{\bf p}(\varepsilon)},
\label{Gk}
\end{equation}
where $\varepsilon({\bf p})$ is the ``bare'' electronic dispersion, while the
total self-energy is given by the additive sum of local $\Sigma (\varepsilon)$, 
determined by  DMFT, and ``external'' $\Sigma_{\bf p}(\varepsilon)$, thus
neglecting any interference between Hubbard and ``external'' interactions. 
This allows us to preserve the structure of self-consistent equations of the
standard DMFT \cite{pruschke,georges96,Vollh10}. Hovewer, there are two major
difference with traditional DMFT. During each DMFT iteration step we
recalculate an ``external'' self-energy $\Sigma_{\bf p}(\varepsilon)$ 
using some approximate scheme, taking into account additional interactions,
and the local Green's function is ``dressed'' by $\Sigma_{\bf p}(\varepsilon)$ 
at each iteration step.

Below for an ``external'' self-energy due to disorder scattering, 
entering DMFT+$\Sigma$ cycle, we use the simplest approximation neglecting
``crossing'' diagrams, i.e. the self-consistent Born approximation, which
in case of Gaussian distribution of site energies takes the (momentum 
independent) form:
\begin{equation}
\Sigma_{\bf p}(\varepsilon)\to\tilde\Sigma(\varepsilon)=\Delta^2\sum_{\bf p}G(\varepsilon,{\bf p}),
\label{BornSigma}
\end{equation}
where $G(\varepsilon,{\bf p})$ is the single-electron Green's function 
(\ref{Gk}) and $\Delta$ is the disorder amplitude.

To solve the effective Anderson impurity problem of DMFT below we use the
numerical renormalization group approach \cite{NRGrev}.

\section{Main results.}

\begin{figure}
\includegraphics[clip=true,width=0.5\textwidth]{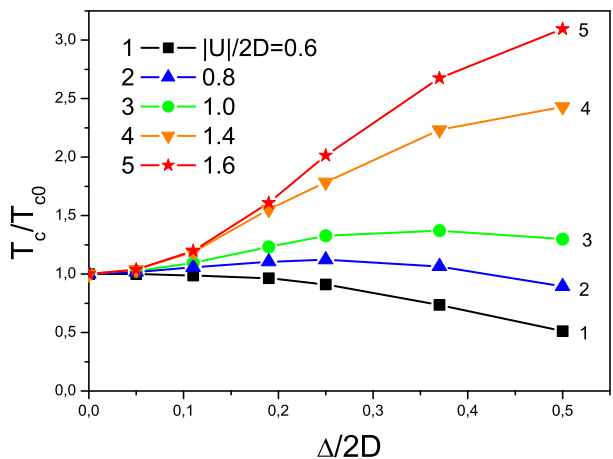}
\caption{Fig. 1. Dependence of superconducting critical temperature on disorder 
for different values of Hubbard attraction.}
\label{fig1}
\end{figure}

In Fig.\ref{fig1} we show the dependence of superconducting transition 
temperature, normalized by the critical temperature in the absence of disorder
($T_{c0}=T_c(\Delta=0)$), for quarter-filled band ($n=0.5$) for different
values of attractive interaction $U$. We can see that in the case of weak coupling
($U/2D\ll1$) disorder somehow suppresses $T_c$ (curve 1). At intermediate 
couplings ($U/2D\sim 1$) weak disorder leads to the growth of $T_c$, 
while the further increase of disorder suppresses the critical temperature
(curves 2 and 3). In the strong coupling region ($U/2D\gg 1$) the growth of disorder
leads to significant increase of the critical temperature (curves 4 and 5). 

\begin{figure}
\includegraphics[clip=true,width=0.5\textwidth]{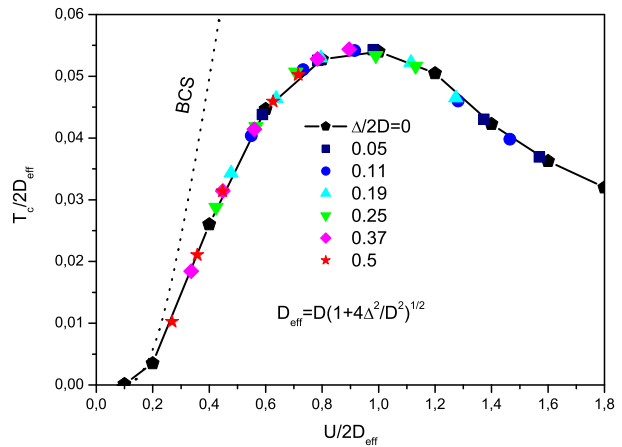}
\caption{Fig. 2. Universal dependence of superconducting critical temperature
on the strength of Hubbard attraction for different values of disorder.}
\label{fig2}
\end{figure}

However, this complicated dependence of superconducting critical temperature
on disorder is easily explained by the conduction band  widening by
growing disorder. In Fig. \ref{fig2} the black curve with pentagonal data points
represents the dependence of critical temperature $T_c/2D$ on attraction
strength $U/2D$ in the absence of disorder ($\Delta =0$) in Nozieres -- 
Schmitt-Rink approximation \cite{JETP14}. The growth of disorder leads to the
effective widening of the conduction band, so that in in our self-consistent Born
approximation for disorder scattering (\ref{BornSigma}) the semi-elliptic
form of the density of states does not change, while the effective half-bandwidth 
grows as \cite{HubDis}:
\begin{equation}
D_{eff}=D\sqrt{1+4\frac{\Delta^2}{D^2}}
\label{Deff}
\end{equation}
The other data points shown in Fig. \ref{fig2} represent the results of our 
calculations in the combined Nozieres --- Schmitt-Rink and DMFT+$\Sigma$
approximations for different values of disorder. We can see that all data points
as expressed via appropriately scaled variables $U/2D_{eff}$ and $T_c/2D_{eff}$
perfectly follow the universal curve, obtained in the absence of disorder.
These results illustrate, at least in approximations used here, the validity of
the generalized Anderson theorem \cite{SCLoc,Genn} (for all couplings, including
the BCS-BEC crossover and strong coupling regions) --- the critical temperature of
superconducting transition (for the case of $s$-wave pairing) is affected by
disorder only through the appropriate change of electron bandwidth (density of 
states).
From Fig. \ref{fig2} we can see, that in the weak coupling region
$U/2D_{eff}\ll 1$ the critical temperature in this approximation is close to 
that obtained in the usual BCS model (dashed curve in Fig. \ref{fig2}). 
For $U/2D_{eff}\sim 1$ the critical temperature $T_c$ reaches the maximum.
For $U/2D_{eff}\gg 1$ it drops with the growth of $U$, showing
$T_c \sim 1/U$ behavior \cite{NS}, as in the strong coupling region $T_c$ is
determined by the condition of Bose--Einstein condensation of Cooper pairs and
hopping  motion of these pairs (via virtual ionization) appears only in the 
second order of perturbation theory being proportional to $t^2/U$ \cite{NS}.

\begin{figure}
\includegraphics[clip=true,width=0.5\textwidth]{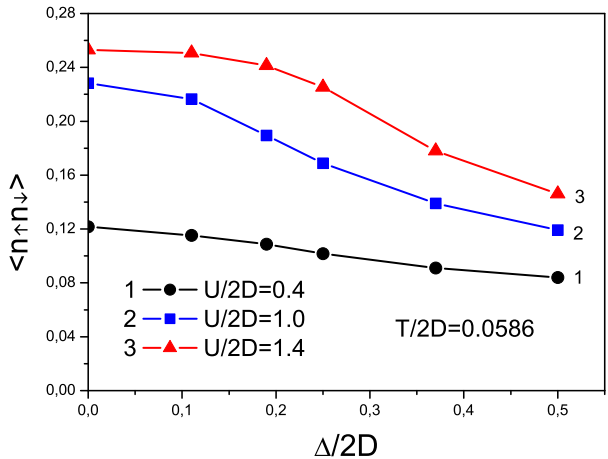}
\caption{Fig. 3. Dependence of the number of local pairs on disorder for
different values of Hubbard attraction.}
\label{fig3}
\end{figure}

\begin{figure}
\includegraphics[clip=true,width=0.5\textwidth]{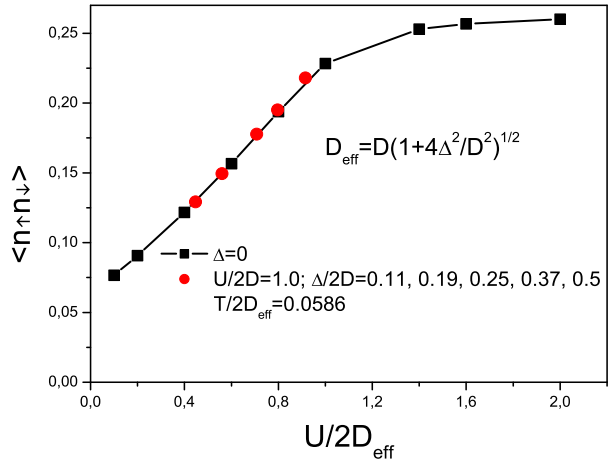}
\caption{Fig. 4. Universal dependence of the number of local pairs on the 
the strength of Hubbard attraction for different values of disorder.}
\label{fig4}
\end{figure}

Band widening due to disorder also leads to the effective suppression of the
number of local pairs (doubly occupied sites). The average number of local
pairs is determined by pair correlation function $<n_{\uparrow}n_{\downarrow}>$, 
which in the absence of disorder grows with the increase of Hubbard attraction
$U$ from $<n_{\uparrow}n_{\downarrow}>=<n_{\uparrow}><n_{\uparrow}>=n^2/4$ 
for $U/2D_{eff}\ll 1$ to $<n_{\uparrow}n_{\downarrow}>=n/2$ for
$U/2D_{eff}\gg 1$, when all electrons are paired. The growth of $D_{eff}$ 
with disorder leads to an effective suppression of the parameter $U/2D_{eff}$ 
and corresponding suppression of the number of doubly occupied sites. 
In Fig. \ref{fig3} we show the disorder dependence of the number of doubly
occupied sites for three different values of Hubbard attraction.
We see that in all cases the growth of disorder suppresses the number of
doubly occupied sites (local pairs).
In fact, similarly to  $T_c$, the change of the number of local pairs with
disorder can be attributed only to the change of the effective bandwidth
of the ``bare'' band (\ref{Deff}) with the growth of disorder. 
In Fig. \ref{fig4} the curve with black squares shows the dependence of the
number of doubly occupied sites on Hubbard attraction for the case of
quarter-filled band ($n=0.5$) in the absence of disorder at temperature
$T/2D=0.0586$. This curve is actually universal --- the dependence of the 
number of local pairs $<n_{\uparrow}n_{\downarrow}>$ on the scaled parameter
$U/2D_{eff}$ with appropriately scaled temperature $T/2D_{eff}=0.0586$ 
in the presence of disorder is given by the same curve, which as shown
by by circles, representing data obtained for five different
disorder levels and shown in Fig. \ref{fig4} for the case of $U/2D=1$.

\section{Conclusion.}

In this paper, using the combined Nozieres -- Schmitt-Rink and DMFT+$\Sigma$
approximations we have investigated the influence of disorder on
superconducting critical temperature and the number of local pairs in
disordered attractive Hubbard model. We have studied the wide range of
attractive couplings $U$, from the weak coupling region of $U/2D_{eff}\ll 1$, 
where normal phase instability and superconductivity is described by BCS model,
to the strong coupling region of $U/2D_{eff}\gg 1$, where superconducting 
transition is related to Bose--Einstein condensation of preformed Cooper pairs,
which appear in the system at temperatures significantly higher, than 
superconducting transition temperature. Disorder can either suppress the
critical temperature $T_c$ in the case of weak coupling, or significantly
increase $T_c$ in the of strong coupling. However, these dependences in fact 
confirm the validity of the generalized Anderson theorem --- all changes of 
superconducting critical temperature can be attributed to general widening of
conduction band by disorder (for the case of $s$-wave pairing, which can only
be realized in the attractive Hubbard model). In the weak coupling region 
transition temperature is well described by BCS model, while in the strong 
coupling region it is determined by the condition of Bose--Einstein condensation 
and drops with the growth of  $|U|$ as $1/|U|$, passing the maximum at
$|U|/2D_{eff}\sim 1$. Similarly, only the band widening by disorder is
responsible for the change of the number of local pairs (doubly occupied sites).
The growth of disorder leads to the effective drop of the ratio
$U/2D_{eff}$ and corresponding drop of the number of local pairs. 

This work is supported by RSF grant No. 14-12-00502.

\newpage

\end{document}